# A QUANTUM ANALOGUE OF PARRONDO'S GAME

**CLEMENT AMPADU**

31 Carrolton Road
Boston, Massachusetts, 02132
U.S.A.
e-mail: drampadu@hotmail.com

**Abstract**

We consider the discrete-time quantum walk in the plane, and present a quantum implementation of Parrondo's game for four players. Physical significance of the game strategies are also discussed.

1.Introduction

The theory of games studies models in which several parties try to maximize their gains by selecting different strategies that are allowed by the rules of a particular game [5]. It has found to have relevance to many disciplines, see for example, the first five references in [2] which covers the social sciences, biology, and economics.

In this paper we present a quantum analogue of Parrondo's game for four players, which we believe is a major step towards providing a framework for applications of quantum walk using multiple coins. Our system involves four players as quantum coin operators to evolve the discrete time quantum walk in the plane. We present a situation where all but one player are unable to manipulate a walk to an extent to win a game using their coins individually. We present a quantum strategy for the players to cooperate by using their quantum coin operations alternatively and emerge as joint winner for situations where it it is conditioned that the winner is decided only after even number of steps of walk evolution. A different joint winning strategy, by using their coins in combination for each step is presented for different situations when winner is not known.

This paper is organized as follows in Section 2 we review basic notions about the discrete time quantum walk in the plane, in Section 3 we give an overview of the Parrondo game in the classical case, in Section 4 we present the quantum analogue of Parrondo's game for four players and the criterion for developing winning strategies, in Section 5 we consider winning strategies for players whom cannot individually win a game using the coin operator accorded to them. Section 6 is devoted to the conclusions, there the significance of our work in applications is briefly discussed

**2. The Discrete-Time Quantum Walk in the Plane**

We recall some facts about the two dimensional quantum walk. In the two-dimensional quantum walk the "coin" degree of freedom is represented by a two-quibit space or coin space, $H_C$, which is spanned by four orthonormal states $\{|L\rangle, |R\rangle, |U\rangle, |D\rangle\}$, where $L, R, U, D$ are associated with the left, right, up and down displacements respectively. The position space, $H_p$, on the other hand is spanned by the set of orthonormal states $\{|x, y\rangle : x, y \in Z, Z \text{ is the integers}\}$. The Hilbert space of the quantum walker is then defined as the tensor product of the coin space $H_C$ and the position space $H_p$. To define the movement of the quantum walker in two dimensions, we recall what happens on one step. We first make superposition on the coin space with the coin operator $U_C$, and move the particle according to the coin state with the translation operator $S$ as follows, $U_W = S \cdot (I \otimes U_C)$ where $I$ is the identity operator in $H_p$, $U_W = S \cdot (I \otimes U_C)$ is the coin operator on the position space and the translation operator $S$ is given by

$$S = \sum_{x,y} \{|x-1, y\rangle\langle x, y| \otimes |L\rangle\langle L| + |x+1, y\rangle\langle x, y| \otimes |R\rangle\langle R| + |x, y-1\rangle\langle x, y| \otimes |D\rangle\langle D| + |x, y+1\rangle\langle x, y| \otimes |U\rangle\langle U|\}$$

The evolution of the two dimensional quantum walk is then defined by $|\Psi(t+1)\rangle = U_w |\Psi(t)\rangle$. By induction on $t$, we can show that evolution of the two dimensional quantum walk in terms of the initial state is given by $|\Psi(t+1)\rangle = U_w^t |\Psi(0)\rangle$. The probability to find the particle at the site $(x, y)$ at time $t$ is given by $P(x, y, t) = \langle x, y | Tr(|\Psi(t)\rangle\langle\Psi(t)|) | x, y \rangle$. Note that choosing $U_c = H**\left(\frac{1}{2}, \frac{1}{2}\right)$ where $H**(p, q)$ is given in Ampadu [1], leads to a standard form of the Discrete Time Quantum Walk in two Dimensions, the Hadamard Walk. Note that $Tr(\cdot)$, is the trace function.

### 3. Overview of Parrondo's Game

Parrondo's game [3, 7] is a 1-player paradoxical game (the player "against the environment"). The player repeatedly chooses from among two strategies $A, B$. Each strategy involves a coin flip; the player adds or subtracts one unit to his capital depending on the flip outcome. The coin is biased, and the bias may depend on the amount of capital accumulated so far. We may choose the bias of both coins to be such that if sequences of strategies AA . . . A or BB . . . B are played then, the capital converges to negative infinity. However, if we switch between the strategies, the capital may converge to positive infinity [5]. Rigorously, we can define Parrondo's game as follows

**Definition 3.1 (Parrondo's Game):** *Parrondo's game is a sequence $\{s(n) \in \{A, B\} : n \in N, N \text{ is the natural numbers}\}$ where $A, B$ are two strategies. Both strategies consist of a coin toss and adding or subtracting one unit of capital to the players account according to the result of the toss. The probability that $A$ wins is $p$; the probability that $B$ wins is $p_0$ if the capital is a multiple of 3 and $p_1$ otherwise.*

### 4. Parrondo's Game Using the Discrete Time Quantum Walk

In this section using the discrete time quantum walk (DTQW) in the plane, we present a novel scheme in which four people can play an analogue of the Parrondo game. Let $P_R$ be the probability that the quantum particle is displaced to the right of the origin, let $P_L$ be the probability that the particle is displaced to the left of the origin, let $P_U$ be the probability that the particle is displaced upward from the origin, and let $P_D$ be the probability that the particle is displaced downward from the origin. Further choose the coin operator $U_C$ as $U_C = B_{\alpha,\beta,\gamma} \otimes B_{\alpha,\beta,\gamma}$, where $B_{\alpha,\beta,\gamma} = \begin{pmatrix} e^{i\alpha}\cos(\beta) & e^{i\gamma}\sin(\beta) \\ e^{-i\gamma}\sin(\beta) & -e^{-i\alpha}\cos(\beta) \end{pmatrix}$ is the coin operator used by Chandrashekar et.al [3] to evolve the DTQW version of the Parrondo game for two players. The explicit form of $U_C = B_{\alpha,\beta,\gamma} \otimes B_{\alpha,\beta,\gamma}$ implies the coin operator is a function of $\alpha, \gamma, \beta, \alpha+\gamma, \alpha-\gamma$. Let us write $U_C = B_{\alpha,\beta,\gamma} \otimes B_{\alpha,\beta,\gamma} \equiv B_{\alpha,\gamma,\beta,\alpha+\gamma,\alpha-\gamma}$. To construct the analogue of the Parrondo game using the DTQW in the plane, we will consider four players $A, B', C', \text{and } D'$ and construct the game as follows.

- Give players $A, B', C', \text{and } D'$ different coin operations say $U_C^A$, $U_C^{B'}$, $U_C^{C'}$ and $U_C^{D'}$ with four nonzero variable parameters each and the common translation operator $S$ defined earlier to evolve the DTQW. To be explicit put $U_C^A = B^A_{0,\gamma,\beta,\alpha+\gamma,\alpha-\gamma}$, $U_C^{B'} = B^{B'}_{\alpha,0,\beta,\alpha+\gamma,\alpha-\gamma}$, $U_C^{C'} = B^{C'}_{\alpha,\gamma,\beta,0,\alpha-\gamma}$ and $U_C^{D'} = B^{D'}_{\alpha,\gamma,\beta,\alpha+\gamma,0}$.

- Let the initial state of the particle at the origin, $(x, y) = (0,0)$, on which the players evolve the walk be given by $|\psi(0)\rangle = \frac{1}{2}(|L\rangle - |R\rangle - |D\rangle + |U\rangle) \otimes |0,0\rangle$

To give criterion for all possible winning combination of players, arbitrarily put $Player\ A, B', C', \text{and } D'$ in quadrants I, II, III and IV in the plane respectively, where the quadrants are determined counterclockwise starting from the positive side of the x- axis, then the interpretation of $P_R$, $P_L$, $P_D$ and $P_U$ implies the winning combination of players and the respective winning criterion are as follows.

- We declare Player $A$ the winner if $P_R > P_L$ and $P_U > P_D$ after $t$ steps of DTQW evolution
- We declare Player $B'$ the winner if $P_L > P_R$ and $P_U > P_D$ after $t$ steps of DTQW evolution
- We declare Player $C'$ the winner if $P_L > P_R$ and $P_D > P_U$ after $t$ steps of DTQW evolution
- We declare Player $D'$ the winner if $P_R > P_L$ and $P_D > P_U$ after $t$ steps of DTQW evolution
- We declare Player $A$ and Player $B'$ joint winners if $P_R = P_L$
- We declare Player $A$ and Player $C'$ joint winners if $P_R = P_L$ and $P_D = P_U$

- *We declare Player $A$ and Player $D'$ joint winners if $P_D = P_U$*

- *We declare Player $B'$ and Player $C'$ joint winners if $P_D = P_U$*

- *We declare Player $B'$ and Player $D'$ joint winners if $P_R = P_L$ and $P_D = P_U$*

- *We declare Player $C'$ and Player $D'$ joint winners if $P_R = P_L$*

- *We declare Players $A, B',$ and $C'$ joint winners if $P_R = P_L$ and $P_D = P_U$*

- *We declare Players $A, B',$ and $D'$ joint winners if $P_R = P_L$ and $P_D = P_U$*

- *We declare Players $B', C',$ and $D'$ joint winners if $P_R = P_L$ and $P_D = P_U$*

- *We declare Players $C', D',$ and $A$ joint winners if $P_R = P_L$ and $P_D = P_U$*

- *All four players can emerge as joint winners if $P_R = P_L = P_u = P_D$.*

Since $U_C = B_{\alpha,\beta,\gamma} \otimes B_{\alpha,\beta,\gamma} \equiv B_{\alpha,\gamma,\beta,\alpha+\gamma,\alpha-\gamma}$, we can write the coin operator on the position space as $U_{\alpha,\gamma,\beta,\alpha+\gamma,\alpha-\gamma} = S \cdot (I \otimes B_{\alpha,\gamma,\beta,\alpha+\gamma,\alpha-\gamma})$, where the translation operator $S$ has the same definition as given earlier on. The action of $U_{\alpha,\gamma,\beta,\alpha+\gamma,\alpha-\gamma} = S \cdot (I \otimes B_{\alpha,\gamma,\beta,\alpha+\gamma,\alpha-\gamma})$ on $|\psi(0)\rangle$ implements one step of the DTQW in the plane, this evolves the quantum particle to $U_{\alpha,\gamma,\beta,\alpha+\gamma,\alpha-\gamma}|\psi(0)\rangle$, explicitly we can write $U_{\alpha,\gamma,\beta,\alpha+\gamma,\alpha-\gamma}|\psi(0)\rangle$ as

$$\frac{1}{2}\left(e^{2i\alpha}\cos^2(\beta) - e^{i(\alpha+\gamma)}\cos(\beta)\sin(\beta) - e^{i(\alpha+\gamma)}\cos(\beta)\sin(\beta) + e^{2i\gamma}\sin^2(\beta)\right)|-1,0\rangle +$$

$$\frac{1}{2}\left(e^{i(\alpha-\gamma)}\cos(\beta)\sin(\beta) + \cos^2(\beta) - \sin^2(\beta) - e^{-i(\alpha-\gamma)}\sin(\beta)\cos(\beta)\right)|1,0\rangle +$$

$$\frac{1}{2}\left(e^{i(\alpha-\gamma)}\cos(\beta)\sin(\beta) - \sin^2(\beta) + \cos^2(\beta) - e^{-i(\alpha-\gamma)}\sin(\beta)\cos(\beta)\right)|0,-1\rangle +$$

$$\frac{1}{2}\left(e^{-2i\alpha}\sin^2(\beta) + e^{-i(\alpha+\gamma)}\cos(\beta)\sin(\beta) + e^{-i(\alpha+\gamma)}\cos(\beta)\sin(\beta) + e^{-2i\alpha}\cos^2(\beta)\right)|0,1\rangle$$

It follows from the expression immediately above for $U_{\alpha,\gamma,\beta,\alpha+\gamma,\alpha-\gamma}|\psi(0)\rangle$ that the position probability distribution after the first step corresponding to the left positions is given by

$\frac{1}{4}\left\{1+\frac{1}{2}\sin^2(2\beta)\cos(2\alpha-2\gamma)-2\sin(2\beta)\cos(\alpha-\gamma)+\frac{1}{2}\sin^2(2\beta)\right\}$, the position probability after the first step corresponding to the right and downward positions is given by

$\frac{1}{4}\left\{1-\frac{1}{2}\sin^2(2\beta)-\frac{1}{2}\sin^2(2\beta)\cos(2\alpha-2\gamma)\right\}$, and the position probability after the first step corresponding to the upward positions is given by

$\frac{1}{4}\left\{1+\frac{1}{2}\sin^2(2\beta)\cos(2\alpha-2\gamma)+2\sin(2\beta)\cos(\alpha-\gamma)+\frac{1}{2}\sin^2(2\beta)\right\}$. Note that for a particle with the initial state $|\psi(0)\rangle = \frac{1}{2}(|L\rangle-|R\rangle-|D\rangle+|U\rangle)\otimes|0,0\rangle$, using an unbiased coin operation, that is, $B_{\alpha,\gamma,\beta,\alpha+\gamma,\alpha-\gamma}$ with $\alpha=\gamma=0$ ($B_{0,0,\beta,0,0} \equiv B_\beta$) can be seen to lead to symmetry of the initial state of the particle. However, from the position probability distributions after the first step, we see that the probability distributions corresponding, for example, to the left and upward positions, would be equal, and lead to what Chandrashekar et.al [2] would call the *left-upward symmetry* if $\alpha=\gamma$, thus in general the operator $B_{\alpha,\gamma,\beta,\alpha+\gamma,\alpha-\gamma}$ as a quantum coin can bias the probability distribution of the quantum walk, despite the symmetry in the initial state of the particle. As Chandrashekar et.al [2] have noted the bias is the key ingredient necessary in developing winning strategies for Parrondo's game using DTQW which we consider in the next section.

Before we concern ourselves with winning strategies we should note that not all of the four players can win a game individually using the quantum coin accorded to them, hence emergence as joint winners is the key with some amount of cooperation. To see why, we recall that the DTQW with $B_{0,0,\beta,0,0} \equiv B_\beta$ on a quantum particle in the initial state $|\psi(0)\rangle = \frac{1}{2}(|L\rangle-|R\rangle-|D\rangle+|U\rangle)\otimes|0,0\rangle$ results in a symmetric distribution. When $\alpha=0$, using Players A coin results in asymmetry with $P_L > P_R$ and $P_D > P_U$ for $-2\pi \leq \gamma \leq -\pi$. When $\gamma=0$, using Players $B'$ coin results in asymmetry with $P_L < P_R$ and

$P_D > P_U$ for $\pi \leq \alpha \leq 2\pi$. When $\alpha + \gamma = 0$, that is, $\alpha = -\gamma$ using players $C'$ coin results in asymmetry with $P_u > P_D$ and $P_R > P_L$ for $-\pi \leq \gamma \leq \dfrac{-\pi}{2}$. When $\alpha - \gamma = 0$, that is, $\alpha = \gamma$, using Players $D'$ coin does not result in asymmetry. As a result we will concentrate only on winning strategies for Parrondo's game using DTQW in the case of Players $A, B'$, and $C'$ in the next section.

## 5. Winning Strategies for Parrondo's game using DTQW.

To device the winning strategies in the case of Players $A, B'$, and $C'$ we will place restrictions on the parameters $\alpha, \gamma, \beta, \alpha + \gamma, and\ \alpha - \gamma$. As we noted in the previous section players $A, B'$, and $C'$ cannot win a game individually, thus it is obvious that we need to consider the cases where the players can emerge as joint winners. To emerge as joint winners some amount of cooperation among the players is necessary, to that end, we will use the levels of cooperation as given by Chadrasherkar et.al [2].

**Definition 5.1 (Cooperation at Level I):** *We say players are cooperating at Level I if each step of the walk is evolved by alternate use of the coins accorded to the players, provided the winner is decided after an even number of steps in the evolution of the DTQW, or by using both of their coins, one after the other for every step of the walk.*

**Definition 5.2 (Cooperation at Level II):** *We say players are cooperating at Level II if they consult among themselves to choose the quantum coin parameters $\alpha, \gamma, \alpha + \gamma, and\ \alpha - \gamma$ (the parameter $\beta$ is common among all the players).*

**Remark 5.3 (Assumptions about levels of cooperation):** *Cooperation at level I is up to the players, since it is necessary for winning jointly. Cooperation at level II is not always up to the players, but this doesn't affect winning jointly.*

We will discuss situations where the players can cooperate at both levels, and when they can only cooperate at level one.

### 5.1. Winning Strategies for Cooperation at Both Levels

In regards to players $A, B'$, and $C'$ we first consider the case where the players are allowed to consult among themselves and the winner is decided after an even number of steps. A simple strategy for players $A$ *and* $B'$ will be to cooperate between themselves for choosing the coin parameter and emerge as joint winners. This can be achieved if both players agree.

1) To use the same value of $\alpha, \gamma, \alpha+\gamma, and\ \alpha-\gamma$ in their respective coins

2) Use their coins every alternative step such that their coins are equally used.

In this case the evolution can be written as,

$$B^A_{0,\gamma,\beta,\alpha+\gamma,\alpha-\gamma} B^{B'}_{\alpha,0,\beta,\alpha+\gamma,\alpha-\gamma} \ldots B^A_{0,\gamma,\beta,\alpha+\gamma,\alpha-\gamma} B^{B'}_{\alpha,0,\beta,\alpha+\gamma,\alpha-\gamma} |\psi(0)\rangle$$

or

$$B^{B'}_{\alpha,0,\beta,\alpha+\gamma,\alpha-\gamma} B^A_{0,\gamma,\beta,\alpha+\gamma,\alpha-\gamma} \ldots B^{B'}_{\alpha,0,\beta,\alpha+\gamma,\alpha-\gamma} B^A_{0,\gamma,\beta,\alpha+\gamma,\alpha-\gamma} |\psi(0)\rangle$$

where $B^A_{0,\gamma,\beta,\alpha+\gamma,\alpha-\gamma} = S\left(B^A_{0,\gamma,\beta,\alpha+\gamma,\alpha-\gamma} \otimes I\right)$, and $B^{B'}_{\alpha,0,\gamma,\beta,\alpha+\gamma,\alpha-\gamma} = S\left(B^{B'}_{\alpha,0,\beta,\alpha+\gamma,\alpha-\gamma} \otimes I\right)$. Note that this strategy is also similar for the other 2-combinations of the three players, and the one 3-combination of the players.

When the players are not allowed to consult each other to choose the coin parameters $\alpha, \gamma, \alpha+\gamma, and\ \alpha-\gamma$, a loosing situation may arise. Since the objective is to win, the players have to be careful about their choice of the coin parameters, to avoid being the loser, that is, either emerge as winner jointly or solely. For example, if we consider the following two-combination of the three players, say $A$ and $B'$, if player $A$ chooses $\gamma = \varepsilon > 0$, where $\varepsilon > 0$ is the smallest possible value allowed in the coin operation, then player $A$ will emerge as solo winner if player $B'$ chooses $\alpha > \varepsilon$, or will emerge as joint winners if player $B'$ chooses $\alpha = \varepsilon$. This strategy also similar to the other 2-combination of the three players. For the one 3-combination of the players, if we set $\alpha + \gamma = \varepsilon = \gamma$, where $\varepsilon > 0$ is the smallest possible value allowed in the coin operation, in the coin operators of the players, then player $A$ can emerge as solo winner over the other players if $\alpha > \varepsilon$. If we set $\alpha = \alpha + \gamma = \varepsilon$ in the coin operators of the players, then player $B'$ can emerge as solo winner over the other players if $\gamma > \varepsilon$. If $\alpha = \gamma = \varepsilon$ in the coin operators of the players, then player $C'$ can emerge as solo winner over the other players, if $\alpha + \gamma > \varepsilon$, it follows that the three players can all be winners simultaneously if we set $\alpha = \gamma = \alpha + \gamma = \varepsilon$ in the coin operators of the players.

**Section 5.2. Winning Strategies for Cooperation at Level One Only**

In the previous section we considered strategies where the players are (i) permitted to consult each other to determine the winner after an even number of steps of walk evolution, and (ii) not permitted to consult each other to determine the winner after an even number of steps of walk evolution. If the number of steps is odd, the player who uses his coin operation one time more than the other player will end being the loser. In this case they can agree upon a new strategy of using both of their coins for each step of the walk such that all the players would have used their coins equally when the winner is decided. If we consider any two-combination of the three players, say players $A$ and $B'$ for example, then the evolution of the walk can be written as

$$B^{AB'}_{\alpha,\gamma,\beta,\alpha+\gamma,\alpha-\gamma} \ldots B^{AB'}_{\alpha,\gamma,\beta,\alpha+\gamma,\alpha-\gamma} B^{AB'}_{\alpha,\gamma,\beta,\alpha+\gamma,\alpha-\gamma} |\psi(0)\rangle \text{ or}$$

$$B^{B'A}_{\alpha,\gamma,\beta,\alpha+\gamma,\alpha-\gamma} \ldots B^{B'A}_{\alpha,\gamma,\beta,\alpha+\gamma,\alpha-\gamma} B^{B'A}_{\alpha,\gamma,\beta,\alpha+\gamma,\alpha-\gamma} |\psi(0)\rangle, \text{ where}$$

$$B^{B'A}_{\alpha,\gamma,\beta,\alpha+\gamma,\alpha-\gamma} = S\left(B^{B'}_{\alpha,0,\beta,\alpha+\gamma,\alpha-\gamma} \otimes I\right)\left(B^{A}_{0,\gamma,\beta,\alpha+\gamma,\alpha-\gamma} \otimes I\right) \equiv S\left(B^{B'}_{\alpha,0,\beta,\alpha+\gamma,\alpha-\gamma} B^{A}_{0,\gamma,\beta,\alpha+\gamma,\alpha-\gamma} \otimes I\right), \text{ and a}$$

similar definition holds for $B^{AB'}_{\alpha,\gamma,\beta,\alpha+\gamma,\alpha-\gamma}$. The evolution of the walk for the one three-combination of the players has a similar description and depends on the player starting the game as in the two-combination case. For example, in the one three-combination of the players, if player A starts the game, then we could write the evolution of walk as

$$B^{AB'C'}_{\alpha,\gamma,\beta,\alpha+\gamma,\alpha-\gamma} \ldots B^{AB'C'}_{\alpha,\gamma,\beta,\alpha+\gamma,\alpha-\gamma} B^{AB'C'}_{\alpha,\gamma,\beta,\alpha+\gamma,\alpha-\gamma} |\psi(0)\rangle, \text{ where}$$

$$B^{AB'C}_{\alpha,\gamma,\beta,\alpha+\gamma,\alpha-\gamma} = S\left(B^{A}_{0,\gamma,\beta,\alpha+\gamma,\alpha-\gamma} \otimes I\right)\left(B^{B'}_{\alpha,0,\beta,\alpha+\gamma,\alpha-\gamma} \otimes I\right)\left(B^{C'}_{\alpha,\gamma,\beta,0,\alpha-\gamma} \otimes I\right)$$
$$\equiv S\left(B^{A}_{0,\gamma,\beta,\alpha+\gamma,\alpha-\gamma} B^{B'}_{\alpha,0,\beta,\alpha+\gamma,\alpha-\gamma} B^{C'}_{\alpha,\gamma,\beta,0,\alpha-\gamma} \otimes I\right)$$

For the example involving the two-combination of the three players $A$ and $B'$, if player $A$ chooses some $-2\pi < \gamma \leq -\pi$, then player $A$ can emerge as solo winner if player $B'$ chooses $\alpha < \gamma$, and player $B'$ can emerge as solo winner if the player chooses $\alpha > \gamma$. If player $A$ starts the game by picking $\gamma = -\pi$, then player $A$ will emerge as joint winner with $B'$ even if $B'$ chooses $\alpha \neq \gamma$. If player $B'$ starts the game and chooses $\alpha = -\pi$, then player $B'$ will emerge as joint winner with $A$ even if player $A$ fixes $-2\pi < \gamma < -\pi$. In the case of the one three-combination of the players, if player $A$ chooses some $-2\pi < \gamma \leq -\pi$, then player $A$ will emerge as joint winner over $B'$ and $C'$ if $B'$ chooses $\alpha > \gamma$ and player $C'$ fixes $-\pi < \gamma < \frac{-\pi}{2}$. If player $B'$ chooses some $\pi < \alpha \leq 2\pi$, then player $B'$ can emerge as solo winner over $A$ and $C'$ if player $A$ chooses $\gamma > \alpha$ and $C'$ fixes $-\pi < \gamma < \frac{-\pi}{2}$. If player $C'$ chooses some $-\pi \leq \gamma < \frac{-\pi}{2}$, then player $C'$ can emerge as solo winner over $A$ and $B'$ if $A$ fixes $-2\pi < \gamma < -\pi$, and $B'$ chooses $\alpha > \gamma$. If player A starts the game and picks $\gamma = -\pi$ will result in a joint winning situation with $B'$ and $C'$, even if player $B'$ fixes $\alpha \neq \gamma$ and $C'$ fixes $-\pi < \gamma < \frac{-\pi}{2}$. If player $B'$ starts the game and picks $\alpha = -\pi$ will result in a joint winning situation with $A$ and $C'$ if $A$ chooses $\gamma \neq \alpha$ and $C'$ fixes $-\pi < \gamma < \frac{-\pi}{2}$. For player $C'$ starting the game and picking $\gamma = -\pi$ will result in joint win $A$ and $B'$ even if $A$ fixes $\alpha \neq \gamma$ and $B'$ fixes $\pi < \alpha < 2\pi$.

**Section 6. Concluding Remarks**

In this paper we have shown how we can implement the parrondo game for four players using notion of the quantum walk in the plane, and presented various strategies for a player to emerge as a solo winner or as a winner jointly with the other players. These strategies have varied consequences in physical situations. For example to arrive at equilibrium, or any non-equilibrium configuration in the probability distribution as required during its application for algorithms or other physical process such as ratchets [2], to understand the anomalous motion of exceptional Brownian particles moving in the opposite direction to the majority [9], to control decoherence [8], to improve quantum algorithms [4], to model lattice gas automata [6], and to understand quantum entanglement [10], just to name a few. In conclusion studying the quantum analogue of the Parrondo game is meaningful work, and can provide useful insights into various quantum information processing tasks and other applications.